\documentclass[10pt,twocolumn]{article}
\newcommand{\settextwidth}[1]{
\setlength{\textwidth}{#1}
\setlength{\oddsidemargin}{\paperwidth}
\addtolength{\oddsidemargin}{-\textwidth}
\setlength{\oddsidemargin}{0.5\oddsidemargin}
\addtolength{\oddsidemargin}{-0.9in}
}
\newcommand{\settextheight}[1]{
\setlength{\textheight}{#1}
\setlength{\headheight}{0mm}
\setlength{\headsep}{0mm}
\setlength{\topmargin}{\paperheight}
\addtolength{\topmargin}{-\textheight}
\setlength{\topmargin}{0.5\topmargin}
\addtolength{\topmargin}{-0.8in}
}
\newcommand{\agent}[1]{\mathit{#1}}
\newcommand{\action}[1]{\mathit{#1}}
\newcommand{\coaction}[1]{\overline{\mathit{#1}}}
\newcommand{\kw}[1]{\mathbf{#1}}
\newcommand{\may}[1]{\langle\langle #1 \rangle\rangle}

\newcommand{\true}{\mathit{true}}
\settextwidth{6.4in}
\settextheight{9in}

\begin{document}
\title{\Large Formal Verification of Quantum Protocols}
\author{\normalsize Rajagopal Nagarajan\thanks{Department of
Computer Science, University of Warwick, Coventry CV4 7AL, UK. Email:
\texttt{biju@dcs.warwick.ac.uk} Phone: +44 247 652 3682 Fax: +44 247
657 3024} \and
\normalsize Simon Gay\thanks{Department of Computing Science, University of
Glasgow, Glasgow G12 8QQ, UK. Email: \texttt{simon@dcs.gla.ac.uk}
Phone: +44 141 330 6035 Fax: +44 141 330 4913}}
\date{}
\maketitle
\pagestyle{empty}
\thispagestyle{empty}
\subsection*{Introduction}
We propose to analyse quantum protocols by applying the
formal verification techniques developed in classical computing for
the analysis of communicating concurrent systems.  Typically, the
first step in formal verification is to define a model of the system
to be analysed, in a well-founded mathematical notation. Experience
has shown that this step in itself is a valuable way of eliminating
ambiguities from an informal description of the system. Next, an
automated analysis tool, based on the same underlying theory, is used
to reason about the system. This might consist either of checking that
the system is behaviourally equivalent to another system which is
viewed as a specification, or of checking that the system satisfies
properties expressed in a separate specification language.

One area of successful application of these techniques is that of
classical security protocols \cite{RyanP:modasp}, exemplified by
Lowe's \cite{LoweG:brefns} discovery and fix of a flaw in the
well-known Needham-Schroeder authentication protocol which had been
proposed several years previously. Secure quantum cryptographic
protocols are also notoriously difficult to design: although protocols
for quantum bit-commitment \cite{BrassardG:quabcs} were believed to be
secure for several years, it has recently been shown not only that
such protocols are insecure, but that secure quantum bit-commitment
is impossible \cite{LoHK:isqbc,MayersD:uncsqb}. Quantum cryptography
is therefore an obvious and interesting target for formal
verification, and provides our first example; we expect the approach
to be transferable to more general quantum information processing scenarios.

Our example is the quantum key distribution protocol proposed by
Bennett and Brassard \cite{BennettCH:quacpd}, commonly referred to as
BB84. We present a model of the protocol in the process calculus CCS
\cite{MilnerR:comc} and the results of some initial analyses using the
Concurrency Workbench of the New Century (CWB-NC)
\cite{CleavelandR:cwbnc}. Similar work could be carried out with other
combinations of modelling language and toolset, such as CSP
\cite{HoareCAR:comsp} and FDR \cite{FSE:fdr}, or Promela and SPIN
\cite{SPIN:spin}.

Proofs of unconditional security of the BB84 protocol exist
\cite{MayersD:uncsqc,LoHK:uncsoq} and we have no reason to doubt their
correctness. Nevertheless, we argue that the modelling/analysis
approach has merit for the study of this and other quantum security
protocols.  Gottesman and Lo \cite{GottesmanD:froqcq} point out
that ``the proof of security of QKD is a fine theoretical result, but
it does not mean that a real QKD system would be secure. Some known
and unknown security loopholes might prove to be fatal. Apparently
minor quirks of a system can provide a lever for an eavesdropper to
break the encryption''. The analysis techniques which we are proposing
can be applied to models at a range of levels of abstraction, from an
idealised description to a concrete implementation.  Moreover, a real
system for security in information processing has components other
than key distribution---authentication or authorisation, for
example. In the future, some of these components may be quantum, but
others could still be classical. We should be able to apply our
methods in a uniform fashion to various components and their
interactions and thus provide certification of complex
systems. Finally, the analysis tools are oriented towards debugging:
if a desired property is not satisfied, then their output enables us
to understand the reason.

\subsection*{Modelling BB84}
We use a version of the BB84 protocol in which Alice reveals the
polarisation basis she used for each photon as soon as the photon is
received and measured by Bob. The CCS model is based on
\emph{processes} and \emph{actions}, both of which may be
parameterised. In this particular model, all parameters are binary
valued.

The quantum communication channel is modelled by a pair of
processes: $\agent{Empty}$ and $\agent{Full}$. The action
$\action{put}(d,b)$ indicates that it is 
possible to send a bit $d$ into the channel; the bit is encoded with respect to
one of two polarisation bases, represented by another binary parameter
$b$. A process which actually sends data into the channel will
contain the complementary action $\coaction{put}$ with a particular
choice of parameters. The dot stands for sequencing, so that the
$\agent{Empty}$ channel becomes $\agent{Full}$ after receiving the data.
\begin{eqnarray*}
\agent{Empty} & = & \action{put}(d,b) . \agent{Full}(d,b)
\end{eqnarray*}
A $\agent{Full}$ channel allows an observer to measure its contents
with respect to a particular basis; the channel then uses the
action $\action{get}$ to release a binary value, and becomes
$\agent{Empty}$. If the basis used for the 
measurement is different from the basis which was originally used to
encode the transmitted bit, then the value released by the channel may
be either 0 or 1. The $+$ operator indicates nondeterminism. Note that
we are only modelling possibilities, not probabilities. Modelling languages and
analysis tools for probabilistic systems are available, for example
PRISM \cite{PRISM:prism}. Probabilistic modelling and reasoning about
quantum protocols is
an area for future work.
\begin{eqnarray*}
\agent{Full}(d,b) &=& \action{measure}(b') . \\
& & \kw{if}\ b'=b \ \kw{then}\ 
\coaction{get}(d) . \agent{Empty}\\
& & \kw{else}\ (\coaction{get}(0)
. \agent{Empty} + \coaction{get}(1) . \agent{Empty})
\end{eqnarray*}
$\agent{Alice}$ and $\agent{Bob}$ interact with the channel via the
actions $\coaction{put}$, $\coaction{measure}$ and $\action{get}$.
The actions $\action{go}$, $\coaction{go}$ are used for
additional synchronisation, so that $\agent{Alice}$ does not 
$\action{choose}$ and $\coaction{put}$ repeatedly before $\agent{Bob}$ has finished processing what he received; this
facilitates later analysis.
\begin{eqnarray*}
\agent{Alice} & = & \action{choose}(x) . (\coaction{put}(x,0)
. \coaction{reveal}(0) . go . \agent{Alice} + \\
              & & \coaction{put}(x,1)
. \coaction{reveal}(1) . go . \agent{Alice}) \\
\agent{Bob} & = & \coaction{measure}(0) . \action{get}(x)
. (\action{reveal}(b) . \kw{if}\ b=0 \\
            & & \ \kw{then}\ \coaction{keep}(x)
. \coaction{go} . \agent{Bob} \ \kw{else}\ \coaction{go}
. \agent{Bob}) \\
& + &  \coaction{measure}(1) . \action{get}(x)
. (\action{reveal}(b) . \kw{if}\ b=1 \\
            & & \ \kw{then}\ \coaction{keep}(x)
. \coaction{go} . \agent{Bob} \ \kw{else}\ \coaction{go}
. \agent{Bob})
\end{eqnarray*}
The complete protocol, without an eavesdropper, consists of the parallel
composition (operator $|$) of $\agent{Alice}$, $\agent{Bob}$, and an
$\agent{Empty}$ channel. The operator $\setminus\{ \action{put},
\action{get},
\action{measure}, \action{go}, \action{reveal} \}$ indicates that these
actions, and their complements, are hidden; they are used for internal
interaction, but are not visible outside. Parallel composition means
that individual processes run independently, only synchronising on
actions and their complements. For example, $\agent{Alice}$'s
$\coaction{put}$ must synchronise with the $\action{put}$ in
$\agent{Empty}$. This means that $\agent{Alice}$ cannot do
$\coaction{reveal}$, synchronising with $\agent{Bob}$, until after
$\agent{Bob}$ has done $\action{get}$; $\agent{Alice}$ has to wait.  
\begin{eqnarray*}
\agent{BB84} & = & (\agent{Alice} | \agent{Bob} | \agent{Empty}) \setminus\\
             &  &\{ \action{put}, \action{get},
\action{measure}, \action{go}, \action{reveal} \}
\end{eqnarray*}
An eavesdropper $\agent{Eve}$ can be modelled similarly, allowing us to
define the attacked protocol $\agent{BB84'}$. This particular eavesdropper
simply guesses a basis, then measures, extracts and returns the sent
bit. More generally, following a standard approach to the analysis of
classical security protocols, we could consider an eavesdropper who
arbitrarily attempts to use any actions with any parameters derivable
from information available to her.
\begin{eqnarray*}
\agent{Eve} & = & \coaction{measure}(0) . \action{get}(x)
. \coaction{put}(x,0) . \agent{Eve} \\
& + & \coaction{measure}(1) . \action{get}(x)
. \coaction{put}(x,1) . \agent{Eve} \\
\agent{BB84'} & = & (\agent{Alice} | \agent{Bob} | \agent{Eve} |
\agent{Empty}) \setminus \\
             &  &\{ \action{put}, \action{get},
\action{measure}, \action{go}, \action{reveal} \}
\end{eqnarray*}
The process $\agent{Spec}$ can be viewed as a specification of the
protocol in the absence of an eavesdropper. $\agent{Spec}$ is a
description of how the protocol should behave: $\agent{Alice}$ chooses
and sends a sequence of bits some of which $\agent{Bob}$ can discard
(when polarisation bases do not match) but whenever he keeps a bit it
is the same as what $\agent{Alice}$ sent. This results in
the generation of a sequence of bits common to both parties---the
\emph{key}.
\begin{eqnarray*}
\agent{Spec} & = & \action{choose}(x) . (\agent{Spec} +
\coaction{keep}(x) . \agent{Spec})
\end{eqnarray*}

\subsection*{Analysing BB84}
Using the tool CWB-NC, we have established that $\agent{BB84}$ is
equivalent to $\agent{Spec}$ and that $\agent{BB84'}$ is not
equivalent to $\agent{Spec}$. ``Equivalent'' refers 
to \emph{trace equivalence}, which means equality of the set of
possible sequences of observable actions. The tool
discovers that $\action{choose}(0)\coaction{keep}(1)$ is a trace of
$\agent{BB84'}$ but not of $\agent{Spec}$. This trace arises from an
execution in which $\agent{Bob}$ measures with the correct basis but
$\agent{Eve}$ has already corrupted the channel by measuring with the
wrong basis. Alternatively, the property
$\may{\action{choose}(0)}\may{\coaction{keep}(1)}\true$ (expressed in
the modal $\mu$-calculus) specifies that a process may choose 0 and
end up keeping 1. CWB-NC establishes that $\agent{BB84}$ does not satisfy
this property, whereas $\agent{BB84'}$ does. This shows the
possibility of interference by $\agent{Eve}$. Note that once the
processes and properties have been defined, CWB-NC carries out
verification automatically and without human intervention.

\subsection*{Conclusions and Future Work}
We have introduced techniques for formally modelling and analysing
quantum protocols. As far as we are aware, this the
first proposal to use formal modelling and analysis in the field of
quantum information processing. As a specific demonstration, we have
modelled components of the BB84 protocol in CCS, and analysed the
model with the CWB-NC tool.

Future work will include the development of a framework based on our
initial investigation, for detailed analysis of quantum information
systems. We are already considering tools that enable us to
incorporate probabilities into our model and we will also include
methods to reason about errors and error-correction. We aim to be able
to generalise the model of the attacker, in order to analyse
collective or coherent attacks, for example. The modelling of
entanglement-based quantum key distribution \cite{EkertAK:quacbb},
other quantum cryptographic protocols and quantum communication
protocols is another goal.

An alternative approach, which we also plan to investigate, is to use
machine-assisted theorem-proving technology to formalise conventional
proofs about quantum systems, such as the unconditional security proofs.

Quantum cryptography is already viable and prototype implementations
are being seriously considered. If verification efforts are begun
early and proceed in tandem with implementations, the resulting
systems are likely to be highly secure.


\small
\begin{thebibliography}{10}
\bibitem{CleavelandR:cwbnc}
{CWB-NC}: \texttt{www.cs.sunysb.edu/\~{}cwb}.

\bibitem{FSE:fdr}
{F}ormal {S}ystems ({E}urope) {L}td: \texttt{www.fsel.com}.

\bibitem{PRISM:prism}
{PRISM}: \texttt{www.cs.bham.ac.uk/\~{}dxp/prism}.

\bibitem{SPIN:spin}
{SPIN}: \texttt{netlib.bell-labs.com/netlib/spin/\\whatispin.html}.

\bibitem{BennettCH:quacpd}
C.~H. Bennett and G.~Brassard.
\newblock Quantum {C}ryptography: {P}ublic-key {D}istribution and {C}oin
  {T}ossing.
\newblock In {\em Proceedings of the IEEE International Conference on
  Computer,\ Systems and Signal Processing, Bangalore, India}, pages 175--179,
  December 1984.

\bibitem{BrassardG:quabcs}
G.~Brassard, C.~Crep\'{e}au, R.~Josza, and D.~Langlois.
\newblock A {Q}uantum {B}it {C}ommitment {S}cheme {P}rovably {U}nbreakable by
  {B}oth {P}arties.
\newblock In {\em Proceedings of the 34th IEEE International Conference on
  Foundations of Computer Science, Palo Alto, California}, pages 362--371. IEEE
  Press, November 1993.

\bibitem{EkertAK:quacbb}
A.~K. Ekert.
\newblock {Q}uantum {C}ryptography based on {B}ell's {T}heorem.
\newblock {\em Physical Review Letters}, 67:661--663, 1991.

\bibitem{GottesmanD:froqcq}
D.~Gottesman and H-K. Lo.
\newblock From {Q}uantum {C}heating to {Q}uantum {S}ecurity.
\newblock {\em Physics Today}, 53(11), November 2000. quant-ph/0111100.

\bibitem{HoareCAR:comsp}
C.~A.~R. Hoare.
\newblock {\em Communicating Sequential Processes}.
\newblock Prentice Hall, 1985.

\bibitem{LoHK:isqbc}
H-K. Lo and H.~F. Chau.
\newblock Is {Q}uantum {B}it {C}ommitment really {P}ossible?
\newblock {\em Physical Review Letters}, 78(17):3410--3413, April 1997. quant-ph/9603004. 

\bibitem{LoHK:uncsoq}
H-K. Lo and H.~F. Chau.
\newblock {U}nconditional {S}ecurity of {Q}uantum {K}ey {D}istribution over {A}rbitrarily {L}ong {D}istances.
\newblock {\em Science}, 283:2050--2056, 26 March 1999. quant-ph/9803006.

\bibitem{LoweG:brefns}
G.~Lowe.
\newblock Breaking and fixing the {N}eedham-{S}chroeder public-key protocol
  using {FDR}.
\newblock {\em Software Concepts and Tools}, 17:93--102, 1996.

\bibitem{MayersD:uncsqb}
D.~Mayers.
\newblock Unconditionally {S}ecure {Q}uantum {B}it {C}ommitment is
  {I}mpossible.
\newblock {\em Physical Review Letters}, 78(17):3414--3417, April 1997. quant-ph/9605044.

\bibitem{MayersD:uncsqc}
D.~Mayers.
\newblock Unconditional {S}ecurity in {Q}uantum {C}ryptography.
\newblock {\em Journal of the ACM}, 48(3):351--406, May 2001.

\bibitem{MilnerR:comc}
R.~Milner.
\newblock {\em Communication and Concurrency}.
\newblock Prentice Hall, 1989.

\bibitem{RyanP:modasp}
P.~Ryan, S.~Schneider, M.~Goldsmith, G.~Lowe, and A.~W.~R. Roscoe.
\newblock {\em Modelling and Analysis of Security Protocols}.
\newblock Addison-Wesley, 2001.

\end{thebibliography}
\end{document}